\documentclass{moriond}

\usepackage{multirow}
\usepackage{csquotes}
\usepackage{orcidlink}
\usepackage{graphicx,wrapfig,lipsum}
\def\be{\begin{equation}}
\def\ee{\end{equation}}
\usepackage{amsmath}
\def\bea{\begin{eqnarray}}
\def\eea{\end{eqnarray}}

\newcommand{\Photo}{\includegraphics[height=35mm]{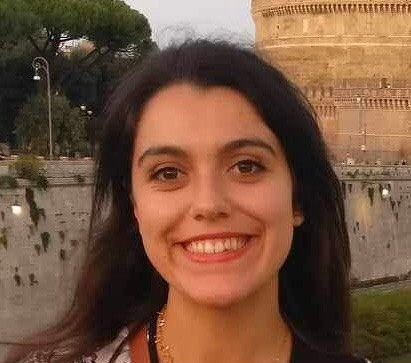}}

\begin{document}
\vspace*{4cm}
\title{Contribution to the 2025 Gravitation session of the 59$^{th}$ Rencontres de Moriond: Limits on the Ejecta Mass During the Search for Kilonovae Associated with Neutron Star-Black Hole
Mergers}

\author{Marion Pillas \orcidlink{0000-0003-3224-2146}}

\address{STAR Institute, Liege University, Allée du Six Août, 19C, B-4000 Liège, Belgium}

\maketitle\abstracts{
This study evaluates ejecta properties from multi-messenger observations to understand the absence of detectable KN associated to the four NSBH candidates from May 2023 to July 2024: we use GW public information and joint observations taken from 05.2023 to 07.2024 (LVK, ATLAS, DECam, GECKO, GOTO, GRANDMA, SAGUARO, TESS, WINTER, ZTF) in the followup of S230518h, GW230529, S230627c and S240422ed. First, our analysis on follow-up observation strategies shows that, on average, more than 50\% of the simulated KNe associated with NSBH mergers reach their peak luminosity around one day after merger in the $g,r,i$- bands, which is not necessarily covered for each NSBH GW candidate. We also analyze the trade-off between observation efficiency and the intrinsic properties of the KN emission, to understand the impact on how these constraints affect our ability to detect the KN, and underlying ejecta properties for each GW candidate. 
}

\section{Introduction} 
GW170817 -- GRB 170817A~\cite{mma170817} was the only confirmed coincidence between a GW signal from binary neutron star (BNS) system and a clear identified kilonova (KN). Together, these gravitational and electromagnetic observations provide complementary information on compact object mergers, including neutron stars (NSs) and black holes (BHs).
Bright electromagnetic emission is not always guaranteed, especially for NSBH mergers and the quantity of the ejecta depends on intrinsic parameters of the binary such as the spins and the mass ratio of the binary system, as well as on the equation of state (EOS) of matter above nuclear saturation density. KNe are widely studied, and numerous models associated with NSBH mergers are now available, with the aim of modeling their production as accurately as possible. We focus in this work on 
Anand et al 2021~\cite{2021NatAs...5...46A} - Bulla 2019~\cite{10.1093/mnras/stz2495} model, later called \textit{An21Bu19} that describes the KN from NSBH with two ejecta components during and after the merger: the \textit{dynamical ejecta}, $m_{dyn}$ that comes from the tidal disruption of the NS and unbound material or \textit{disk wind ejecta} from the post-merger accretion disk (referred as $m_{wind}$). The fourth LIGO/Virgo/KAGRA observing run O4 is ongoing and up to 2024 July 24, one NSBH merger has been confirmed, GW230529 and three NSBH merger candidates were announced: S230518h, S230627c, S240422ed, the latter now being considered as marginal. None of these candidates led to the discovery of a confirmed EM counterpart despite follow-up observations. 
In this work, we first assess the impact and importance of observations occurring at the time of predicted brightness peak, and how it affects the detection. Then, we evaluate the range of ejecta masses that informs the r-process synthesis and KN emission, based of GW information and finally we compare results with both observations of GW and optical data to constrain extrinsic and intrinsic parameters of the NSBH candidates and their astrophysical origin.
\section{Spatial and temporal observation strategy}
\label{sec:observation-strategy}
\begin{figure}[t]
    \centering
    \includegraphics[width=\linewidth]{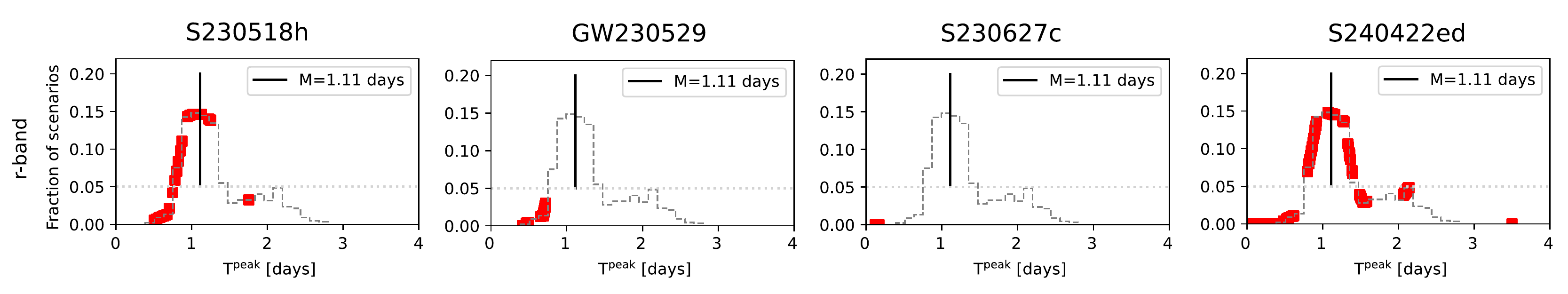}
    \vspace*{-1cm}
    \caption{Comparison between the peak time luminosity of our simulated KN dataset in $r$-band and the time of optical observations for the four NSBH candidates. The dashed gray line represents the distribution of peak time of the simulated dataset. The solid vertical black line represents the median of the peak time distribution considering only bins containing more than 5\% of the distribution. Observations are shown in red.}
    \label{fig:peak-time}
\end{figure}

We explore the optimization of the detection of KNe. In this section, we summarize the total spatial coverage for each candidate and we then focuses on the temporal efficiency of observations to discover KNe. We assume that the observations should seek to cover the \textquote{peak time} \textit{e.g.} the time that corresponds to the peak luminosity of a scenario, at least once in order to optimize the likelihood of counterpart detection. To do this, we compare the time of the observations taken by the community for each NSBH alert to the peak magnitude of simulated KNe light curves computed from \textit{An21Bu19} model.
Each simulated KN defines a different possible \textquote{scenario} that corresponds to different properties of the dynamical and disk wind ejecta between 0.01 and 0.09$~M_\odot$ and viewing angle between 0 and 90$^\circ$. Fig.~\ref{fig:peak-time} shows histograms of peak time for simulated KNe from NSBH mergers for $r$-band in dashed gray line. In Pillas et al, 2025 \cite{pillas2025limitsejectamasssearch} we show similar figures for other filters. As a summary: \textbf{S230518h --} In the optical range, there was a total of 81\% coverage of the most recently updated sky localization for a magnitude limit ranging from 14.5 to 23.3 in various bands. In $r$-band, the beginning of the histogram is well-covered (70\%), with observations occurring at the maximum of the peak time distribution. Observations using $tess$ filter covered almost 100\% of the peak time of the KNe population (see Appendix in Pillas et al, 2025 \cite{pillas2025limitsejectamasssearch}). \textbf{GW230529 --} This significant event has been infrequently observed by the community due to the very poorly constrained sky localization. In the optical range, there was a total of 37\% coverage of the most recently updated sky localization for a magnitude limit ranging from 13.2 to 21.7 in various bands. Observations of GW230529 from 0 to 6 days in $r$-band cover only 2\% of the peak time of the synthetic KNe population. Finally, observations in $L$ covered 44\% of the peak time of synthetic KNe population (see Appendix in Pillas et al, 2025 \cite{pillas2025limitsejectamasssearch}). \textbf{S230627c --} In the optical range, there was a total of 96\% coverage of the most recently updated sky localization for a magnitude limit ranging from 16.3 to 21.3 in various bands. Observations of S230627c did not cover the peak time of our KNe population in $r$-band, as they were performed prior to 0.5 days post T$_\mathrm{GW}$. Observations with the most frequently used filter, $R$-band, covered 26\% of the peak time of the KNe population (see Appendix in Pillas et al, 2025 \cite{pillas2025limitsejectamasssearch}). \textbf{S240422ed --} Initially classified as $>$~99\% likely an NSBH merger, this candidate was, two months later, reclassified as 93\% likely non-astrophysical. Considering all observations in the optical, there was a total of $\sim$99.7\% coverage of the most recently updated sky localization for a magnitude limit ranging from 14.1 to 23.5 in various bands between 0 and 6 days post-T$_\mathrm{GW}$. Observations of S240422ed covered 82\% of the peak time of our KNe population in the $r$-band.

In summary, observations at the predicted brightness peak of the KN are crucial for maximizing detection. Our study highlights not only the need for prompt imaging of NSBH mergers but also emphasizes the importance of 1-day post-merger observations in the UVOIR bands. In partucular, observations taken from 0.9 to 1.4\,days post T$_\mathrm{GW}$ in $g,r,i$-bands can cover the peak time of $>$ 51\% of the synthetic KNe from our population. Although the community has effectively implemented prompt observation strategies, 1-day post-T$_\mathrm{GW}$ observations were not performed in all cases. We also advocate a more \textquote{flexible} approach for near-infrared and infrared observations as the peak luminosity distribution of KNe is relatively flat. 

\section{Ejecta computation and Constraints on kilonova emission}
\label{sec:ejecta}
\begin{figure}
    \centering
    \includegraphics[width=\linewidth]{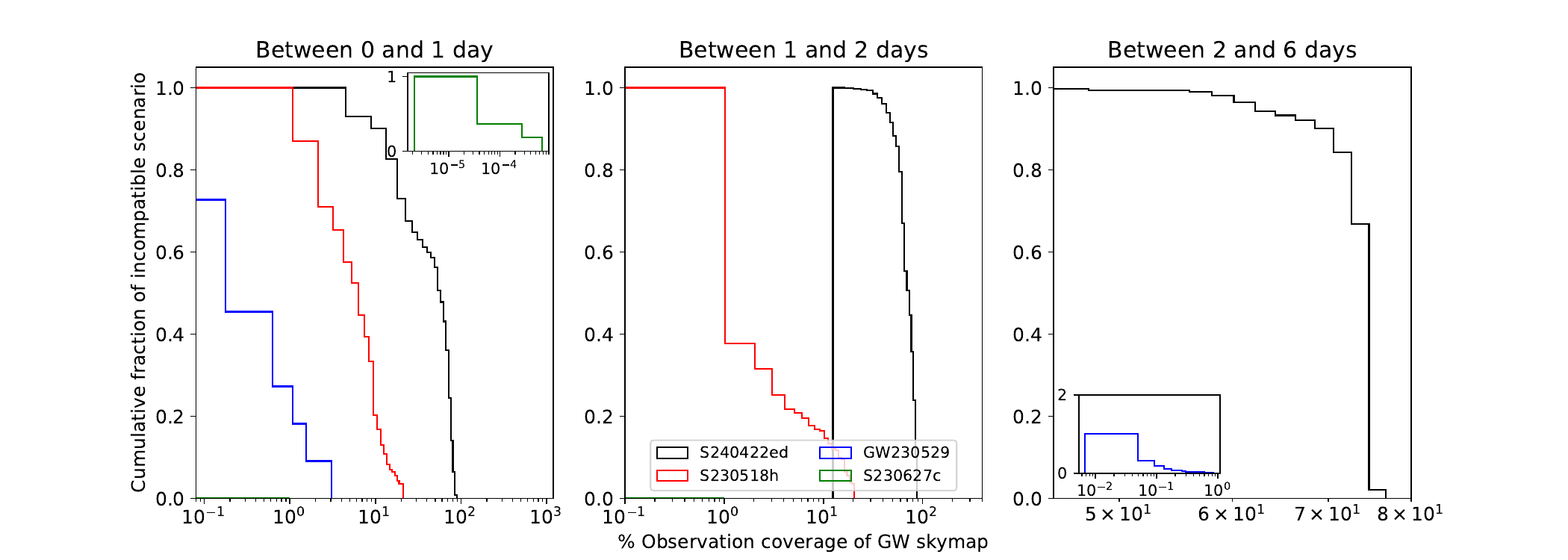}
    \vspace*{-0.8cm}
    \caption{Cumulative histograms showing the fraction of KN scenario that are incompatible with optical observations as a function of the GW skymap coverage that rules them out, for each NSBH candidates (colored line).}
    \label{fig:cumulative-hist-prob-coverage}
\end{figure}

In our study, we use fitting formulae calibrated to the result of merger and post-merger simulations to estimate $m_{dyn}$ and $m_{wind}$. The relevant formulas can be found in Kruger et al. 2020~\cite{2020PhRvD.101j3002K} (Eq.~9), Foucart et al. 2018~\cite{2018PhRvD..98h1501F} (Eq.~4) and Raaijmakers et al. 2021~\cite{2021ApJ...922..269R} (Eq.~12). 
We then constrain the ejecta mass ($m_{dyn}$, $m_{wind}$) of S230518h, S230627c and S240422ed considering each event candidate's classification (\textquote{p-astro}) probabilities contained in the LVK alerts. In case of the detection of an event, thes CBC searches report publicly their source classifications in \href{https://gracedb.ligo.org/}{Gracedb}. 
The compact binary coalescence search pipeline PyCBC Live's method~\cite{Villa-Ortega:2022qdo} provides a way to use public candidate information to constrain source properties and provide an upper limit on the ejected matter during and after the merger. This approach creates a deterministic mapping between the source-frame's chirp mass and the "p-astro" assuming an astrophysical origin of the event, which can be used to constrain the ejecta. Thus, we can provide broad ejecta upper limit boundaries which corresponds to the optimistic case testing several EOS and spin scenario. This gives the following upper limits for each NSBH candidate: \textbf{S230518h} -- $m_{dyn}~<~0.08~\mathrm{M}_\odot$ and $m_{wind}~<~0.04~M_\odot$;
\textbf{GW250529} -- $m_{dyn},~ m_{wind}~\leq~0.01~M_\odot$;
\textbf{S230627c} -- $m_{dyn},~ m_{wind}~\leq~0.01~M_\odot$;
\textbf{S230627c} -- given the low significance of this candidate, we select all the synthetic light curves of the model grid in order to remain as agnostic as possible. Finally, we select synthetic light curves consistent with these upper limits.
\begin{wrapfigure}{r}{8cm}
     \includegraphics[width=8cm]{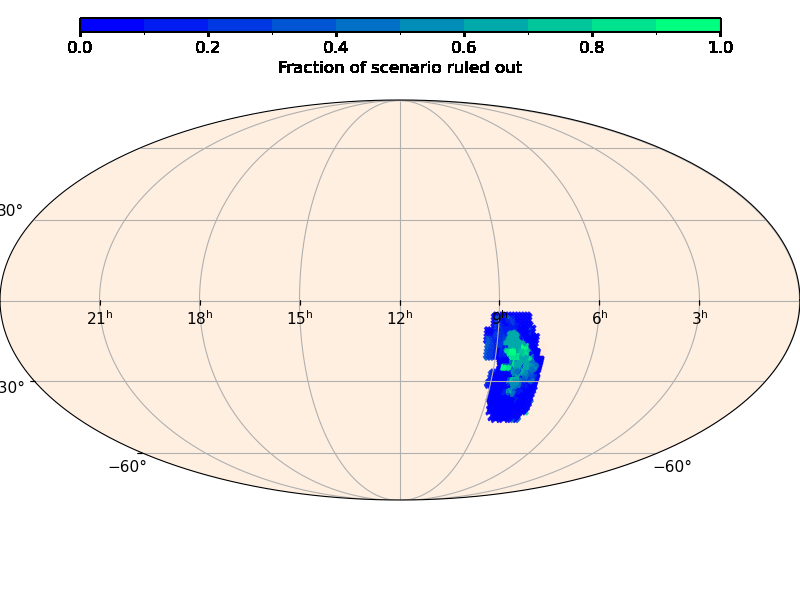}
     \vspace*{-1.6cm}
    \caption{S240422ed 2-D histogram over the sky of the fraction of scenarios (in any filter) that rule out the presence of the KN in the observations.}
    \label{fig:skymap-mag-S240422ed}
\end{wrapfigure} Then, to test the models against the observations, we proceed as follows: for each field observed by an optical telescope, we extract the corresponding pixels and their associated distances in the GW HEALPix skymap. We compute the apparent magnitude of the selected synthetic KN light curves at these distances. We can hence compare the brightness of the selected simulated KNe with the limiting magnitude of the fields at the time of the observation. Then, if a simulated KN luminosity in the detector frame is brighter than the limiting magnitude of the observation, we conclude that the simulated KN emission and properties are not compatible with the observations. We scan the GW skymap and for each pixel we compute the fraction of KN scenario ruled out by observations on top of the pixel (Eq.~1 of Pillas et al, 2025~\cite{pillas2025limitsejectamasssearch}). Figure~\ref{fig:skymap-mag-S240422ed} illustrates the it over the S240422ed GW skymap between 0 and 1 day. Here we rule out the presence of a KN (i.e., 70\% of our synthetic KN population is inconsistent with the observed data) over an area of 43 deg² within the 90\% credible region within the first day post T$_\mathrm{GW}$, 153 deg² within the 90\% S240422ed credible region between 1 and 2 days post T$_\mathrm{GW}$, and 178 deg² thereafter.
Figure~\ref{fig:cumulative-hist-prob-coverage} presents our final result the fraction of incompatible KN scenarios as a function of the percentage of skymap coverage (weighted by the pixel's probability) that rules them out. We summarize our results event by event below: \textbf{S230518h} -- According to Fig.~\ref{fig:cumulative-hist-prob-coverage}, a fraction of 34\% of KNe are incompatible with observations covering more than 8\% of S230518h sky localization region between 0 and 1 day. In details, it has not been possible to observe KNe emitted from an on-axis collision up to a viewing angle of $\theta = 25^\circ$, assuming a minimum confidence of 8\% for the presence of the source in this region. Finally low-mass ejecta ($m_{dyn},m_{wind} \leq 0.03 M_\odot$) seem to be favored. \textbf{GW250529 \& S230627c} -- Overall, observations that assess no KNe compatible with GW cover less than 3\% of GW230529 and S230627c skymaps as shown in Fig.~\ref{fig:cumulative-hist-prob-coverage}, therefore, we cannot place constraints on the viewing angle and ejecta properties. \textbf{S240422ed} -- Fig.~\ref{fig:cumulative-hist-prob-coverage} shows that 59\%, 93\%, $\sim$100\% of the simulated KNe are incompatible with observations covering more than 45\% of S240422ed skymap in [0,1], [1,2] and [2,6] days respectively.

\section{Conclusion}
\label{sec:discussion}
By projecting the subset of synthetic KNe onto 3D skymaps provided by LVK alerts, we measured compatibility with non-detections and observational upper limits. As a result, regarding the confirmed event GW230529 and S230627c candidate, we cannot place constraints on the ejecta or viewing angle of the source, due to telescope observational efficiency. For S230518h, low-mass ejecta and viewing angle above 25$^\circ$ are favored. Eventually, for the low-significance candidate S240422ed, observations ruled out the presence of a KN. This absence of KN is not consistent with the PyCBC Live p$_\mathrm{astro}$ ($p_{BNS}~=~0.7$ when the $p_{astro}$ is rescaled to 1) as a null ejecta would correspond to a probability of BNS classification at least lower than $\sim$ 0.3. Our result seems to be in favor of the non-astrophysical origin of the candidate, which would be consistent with the downgrade of S240422ed significance by LVK collaboration.

\section*{References}
\bibliography{moriond}

\begin{thebibliography}{1}

\bibitem{mma170817}
{LIGO Scientific Collaboration} et~al.
\newblock {\em APJL}, 848(2):L12, October 2017.

\bibitem{2021NatAs...5...46A}
Shreya~Anand et~al.
\newblock {\em Nature Astronomy}, 5:46--53, January 2021.

\bibitem{10.1093/mnras/stz2495}
M~Bulla.
\newblock {\em Monthly Notices of the Royal Astronomical Society}, 489(4):5037--5045, 09 2019.

\bibitem{pillas2025limitsejectamasssearch}
M.~Pillas and et~al.
\newblock https://arxiv.org/abs/2503.15422, 2025.

\bibitem{2020PhRvD.101j3002K}
Christian J~Kr{\"u}ger et~al.
\newblock {\em PRD}, 101(10):103002, May 2020.

\bibitem{2018PhRvD..98h1501F}
Francois~Foucart et~al.
\newblock {\em PRD}, 98(8):081501, October 2018.

\bibitem{2021ApJ...922..269R}
Geert~Raaijmakers et~al.
\newblock {\em APJ}, 922(2):269, December 2021.

\bibitem{Villa-Ortega:2022qdo}
Ver\'onica Villa-Ortega et~al.
\newblock {\em Mon. Not. Roy. Astron. Soc.}, 515(4):5718--5729, 2022.

\end{thebibliography}


\end{document}